\documentclass{aastex62}

\published{{2023 December 20}}
\usepackage{epstopdf}
\shorttitle{Flare Productivity}
\shortauthors{Dhakal et al.}
\begin{document}

\title{What are the Causes of Super Activity of Solar Active Regions?}

\correspondingauthor{Suman Dhakal}
\email{sdhakal2@gmu.edu}

\author{Suman K. Dhakal}

\author{Jie Zhang}
\affiliation{Department of Physics and Astronomy, George Mason University, Fairfax, VA 22030, USA}
\

\begin{abstract}
Flare productivity varies among solar active regions (ARs). This study analyzed 20 ARs of contrasting sunspot areas and flare productivities to understand the super flare productivity of certain ARs. We used the flare index (FI) as an indicator of flare activity. We examined the pattern of morphological evolution of magnetic features. Further, we derived a set of magnetic feature parameters to quantitatively characterize ARs. Our study found that the correlation coefficient is the highest ($r$ = 0.78) between FI and the length of the strong gradient polarity inversion line (SgPIL), while the coefficient is the lowest ($r$ = 0.14) between FI and the total unsigned magnetic flux. For the selected ARs, this study also found that the super flare productive ARs have SgPILs ($R$ value) longer (greater) than 50 Mm (4.5). These results suggest that flare productivity is mainly controlled by the size of the subregion that comprises close interaction of opposite magnetic polarities and is weakly correlated with the size of the whole ARs. Further, even though magnetic flux emergence is important, this study shows that it alone is insufficient to increase flare productivity. New emergence can drive either the interaction of like or opposite magnetic polarities of nonconjugate pairs (i.e., polarities not from the same bipole).  In the former case, the magnetic configuration remains simple, and flare productivity would be low. In the latter case, the convergence of opposite magnetic fluxes of nonconjugate pairs results in a magnetic configuration with long SgPIL and an increase in flare productivity.

\end{abstract}


\section{Introduction} 
\label{sec:intro}
Solar flares are observed as an intense and impulsive release of electromagnetic radiation from localized areas of the Sun. They can be seen over the entire electromagnetic spectrum, ranging from radio to $\gamma$-rays.  Usually, soft X-ray (SXR) observations from Geostationary Operational Environmental satellites (GOES) are used to identify and classify flares (see \citealt{Fletcher_etal_2011} and references therein). Flares are caused by the release of magnetic energy stored in the Sun's corona through the process of magnetic reconnection (see \citealt{Shibata_and_Magara_2011} and references therein).  Despite numerous observational and modeling studies on flares, flare prediction is still challenging.

Active regions (ARs), regions of strong concentration of magnetic fluxes of both positive and negative polarities, are the main sources of flares. The life cycle of an AR begins with the emergence of opposite magnetic fluxes, known as the emergence phase, on the photosphere. Newly emerging positive (negative) magnetic fluxes merge together to form a region of strong concentration of positive (negative) polarity. The next phase is the decay phase, where the area of concentrated magnetic fluxes starts to break, diffuse, and dissipate from the photosphere. It is important to note that an AR can have multiple episodes of magnetic flux emergence, where new magnetic fluxes emerge within or near the existing magnetic fluxes (see \citealt{van_Driel_and_Green_2015}; \citealt{Cheung_etal_2017} for review).

The level of flare activity can change over the life cycle of an AR. It could be higher during the emergence phase than in the decay phase (e.g. \citealt{Choudhary_etal_2013}). In some cases, flare activity can increase during the decay phase of ARs (e.g. \citealt{Patty_and_Hagyard_1986}). One of the most debated issues regarding the origin of flares is whether the intensity and frequency of flares depend on the magnetic structures of emergence or the surface evolution of magnetic fluxes after the emergence. \citet{Schrijver_2007} studied over 2500 AR magnetograms and found that the X- and M-class flares are associated with distinct strong-gradient polarity inversion lines (SgPILs). They suggest that SgPILs are formed due to the emergence of current-carrying magnetic fields and such emergence is responsible for the intense flare. However, some recent studies (e.g., \citealt{Chintzoglou_etal_2019}; \citealt{Liu_etal_2021}) suggest that long SgPILs in ARs are formed due to the interaction between opposite magnetic fluxes of nonconjugate polarity pairs (i.e, positive and negative magnetic poles that do not emerge together as a bipole) and such interaction leads to intense flares. Based on such observations, they argue that shearing motion and flux cancellation are the main drivers of flares.

 Several observational magnetic properties are associated with higher flare productivity. Past studies have shown that flare productivity is correlated with complex polarity patterns like $\delta$-configuration (e.g., \citealt{Shi_and_Wang_1994}), anti-Hale magnetic configuration (e.g., \citealt{Tian_etal_2002}), and AR's size (e.g., \citealt{Yang_etal_2017}). Besides direct observational features, many magnetic feature parameters are used to understand the flare productivity of ARs. These parameters, such as total magnetic flux content, magnetic shear, current helicity, and $R$ value (e.g.,  \citealt{Leka_and_Barnes_2003a}; \citealt{Schrijver_2007}), are calculated using photospheric magnetogram data. Such parameters are used as a proxy to understand the coronal condition of an AR and to understand flare productivity or for flare prediction (e.g., \citealt{Bobra_and_Couvidat_2015}; \citealt{Li_etal_2021}; \citealt{Ran_etal_2022}). Therefore, it is important to understand how these parameters change with the evolution of ARs and which parameters are important to reflect flare productivity.

In this study, we focus on ARs with super flare activity and intend to identify the physical evolutional processes as well as feature parameters of photospheric magnetic fields that are important to the super activity of ARs through a comparative study of 20 ARs of varying sizes and flare productivities. We use the flare index (FI; see Section~\ref{instru}) to compare the flare productivity of ARs. These ARs are selected from the following four contrasting groups; (i) ARs with small sunspot areas ($<$~700 MSH; millionths of solar hemisphere) and small FIs ($<$~20.0) ,(ii) ARs with large sunspot areas ($>$~700 MSH) and small FIs ($<$~20.0), (iii) ARs with large sunspot areas ($>$~700 MSH) and large FIs ($\ge$~20.0), and (iv) ARs with small sunspot areas  ($<$~700 MSH) and large FIs ($\ge$~20.0). Note that the ``large'' and ``small'' are relative terms, and the choice of thresholds is discussed in Section~\ref{instru}. Further, we calculate six magnetic feature parameters that are important and relevant and correlate them with the FI of ARs. We also inspect the evolution of the magnetic configuration of these ARs during their entire transit across the front disk of the Sun. Our study suggests that the persistent interaction of nonconjugate pairs is the dominant cause of super flare productivity. The paper is structured as follows. The data and methodology are described in Section~\ref{instru}. the evolutions of ARs are compared in Section~\ref{AR_evolution}. Magnetic feature parameters are analyzed in Section~\ref{flare_driver}, and results and discussion are presented in Section~\ref{DC}.

\section{\textbf{Observation and Data Analysis}}
\label{instru}
In this study, we primarily used the data from the Helioseismic and Magnetic Imager (HMI; \citealt{Schou_etal_2012}) on board the Solar Dynamics Observatory (SDO;~\citealt{Pesnell_etal_2012}). We used HMI line-of-sight magnetogram ($B_{los}$; with a cadence of 45 s) data to study the evolution of the ARs. Additionally, we used the processed Spaceweather HMI Active Region Patch (SHARPs; \citealt{Bobra_etal_2014}; \citealt{Hoeksema_etal_2014}) vector magnetogram data series, with a cadence of 12 minutes, for the calculation of magnetic feature parameters. The SHARP data are in Lambert cylindrical equal-area (CEA) projection, where all pixels have equal area. As the analysis of each AR spans many days, the physical parameters of interest are calculated and analyzed at an hourly cadence from a cutout of vector magnetograms containing the AR. The size of the cutouts was kept the same in such a way that they encompassed all the major magnetic polarities during the observational period of ARs.

For the comparative study of ARs, we selected the ARs from an AR flare catalog. The catalog was generated using the data from NOAA's Solar Region Summary (SRS)\footnote{https://www.swpc.noaa.gov/products/solar-region-summary}, GOES flare list\footnote{https://hesperia.gsfc.nasa.gov/goes/goes\_event\_listings/}, and Solarsoft latest events archive\footnote{https://www.lmsal.com/solarsoft/latest\_events/}. It has information on the location of the sunspot, sunspot area, magnetic configuration, and flares during the front-disk passage of each AR. Furthermore, we calculated and assigned an average flare index (FI) to each AR to characterize its level of flare productivity. We calculated the average FI as:

\begin{equation}
$$$FI$  = $\frac{(100\sum\limits_{i=1}^{N_{x}} X_{i} +10\sum\limits_{i=1}^{N_{m}}M_{i}+1\sum\limits_{i=1}^{N_{c}}C_{i})}{N_{D}}$$$
\end{equation}

$C_{i}$, $M_{i}$, and $X_{i}$ are the magnitudes of individual C-, M-, and X-class flares, respectively. For instance, $C_{i}$ = 5.3 for a C5.3 flare. $N_{c}$, $N_{m}$, and $N_{x}$ are the number of C-, M-, and X-class flares that occurred in that AR over the entire period of tracking. $N_{D}$ is the total number of days an AR was tracked on the front disk. Here, SXR flares of classes C, M, and X are weighted as 1, 10, and 100, respectively, in units of C1.0 flare or 10$^{-6}$ W m$^{-2}$ (e.g., \citealt{Antalova_1996}; \citealt{Abramenko_2005}). The usage of FI makes it straightforward to compare the relative flare productivity of ARs.

\begin{figure*}[!ht]
\centering
\includegraphics[width=.4\textwidth,clip=]{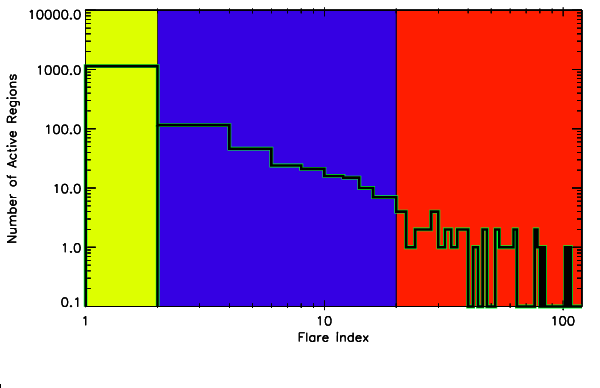}
\caption{Distribution of FI of the ARs identified between 2010 and 2015. The region with the yellow background shows the ARs with low flare activity. The region with the blue background shows the ARs with moderately flare activity. The region with the red background shows the ARs with high flare activity.}
\label{fig:hist}
\end{figure*}

During the six-year span from 2010 to 2015 inclusive, 1437 ARs are registered to appear on the front disk of the Sun, either by rotation from the back side or through new emergence on the front side. Figure \ref{fig:hist} shows the histogram distribution of the number of ARs with respect to FI. The average FI for all these ARs is 2.78, and the median FI is 0.30. For the convenience of a contrasting study pursued in this article, we classified these ARs into three activity categories: (1) low-active ARs (LAARs) of FI $<$ 2.0, (2) moderately-active ARs (MAARs) of 2.0 $\le$ FI $<$ 20.0, and (3) high or super-active ARs (SAARs) defined by FI $\ge$ 20.0. The LAARs have a FI lower than the average value but consist of the majority of ARs ($\sim$79$\%$). About 18$\%$ of the ARs fall into the MAAR category. On the other hand, only 3$\%$ (or 39 out of 1437) of the ARs belong to the category of SAARs. Note that the usage of 20.0 to separate SAARs and MAARs is rather arbitrary.  An AR of FI of 20.0 means that on average the AR produces two M-class flares every day during its many-day transit over the Sun's front disk. Further, the threshold value is about 7 times the average FI and about 67 times the median FI. We have chosen a rather high threshold (FI $\ge$ 20) to define SAARs with the objective of identifying the dominant physical processes responsible for the super flare productivity of ARs.

\begin{figure*}[!ht]
\centering
\includegraphics[width=.4\textwidth,clip=]{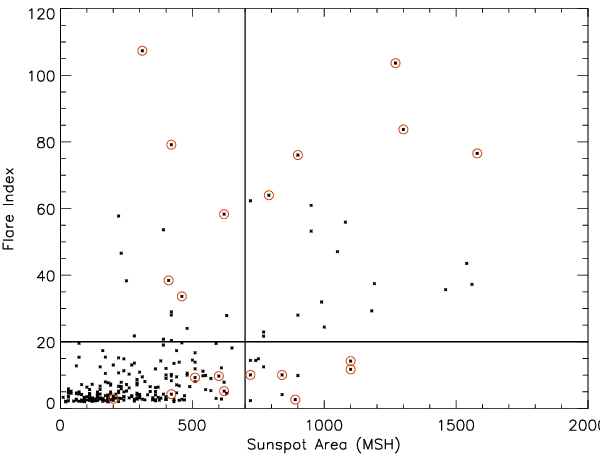}
\caption{Distribution of FI with respect to the sunspot area. The vertical line separates the ARs into groups of large and small sunspot areas. The horizontal line separates the ARs into groups of super and moderately flare-productive ARs. The ARs selected for the comparative study are circled in red. Note that LAARs are not shown.}
\label{fig:fi_area}
\end{figure*}

To select ARs for a meaningful comparison, they were sorted using the following selection criteria: (a) FI $\ge$ 2 and (b) front-disk visibility for more than 7 days. Among 1437 ARs, only 258 ARs met the selection criteria. Figure \ref{fig:fi_area} shows the scatter plot of FI versus the sunspot area of these ARs. The choice of the 7 day threshold was to have long-enough observation of ARs to differentiate the changes in flare productivity with different evolutionary phases, possibly having both emerging and decaying phases of ARs. The selection criteria excluded the ARs emerging near the western limb of the Sun and decayed ARs. Note that AR 12192 is not shown in Figure \ref{fig:fi_area} due to its exceptionally large area (2750 MSH) and FI (164). For the approach of a comparative study of the flare productivity of ARs, we divided the ARs into the following four groups (see Figure \ref{fig:fi_area}):

\begin{enumerate}
\item {\bf \textit {group I:}} ARs with relatively small sunspot areas ($< $700 MSH) and relatively small FIs ($< $20). Most of the ARs belong to this group; in total there are 209 ARs in this group.
\item {\bf \textit {group II:}} ARs with large sunspot areas ($\ge$ 700 MSH) and relatively small FIs ($<$ 20). Few ARs belong to this group, i.e., a total of 12 ARs.
\item {\bf \textit {group III:}} ARs with large sunspot areas ($\ge$ 700 MSH) and large FIs ($\ge$ 20). There are 21 ARs belonging to this group. 
\item {\bf \textit {group IV:}} ARs with relatively small sunspot areas ($<$ 700 MSH) and large FIs ($\ge$ 20). Fewer ARs belong to this group; in total there are 16 such ARs.
\end{enumerate}

The boundaries separating different groups are fixed arbitrarily to facilitate the grouping of ARs for this particular study.  ARs above (below) the horizontal line (see the right panel of Figure ~\ref{fig:fi_area}) belong to SAARs (MAARs). ARs on the right side of the vertical lines are large and on the left side are small (see Figure~\ref{fig:fi_area}). It is important to note that the usage of ``large" and ``small" in this study is relative. Comparing the number of ARs on both sides, it is clear that large ARs are rare. The correlation between the sunspot area and FI of ARs satisfying our selection criteria is 0.67. Such a moderate correlation suggests that, generally, flare productivity increases with the sunspot size of ARs. However, ARs in  group II (ARs with large sunspot area and small FI) and group IV (small sunspot area and large FI) deviate from the expectation that flare productivity depends on the sunspot size of the ARs. Analysis of such deviations would help to understand the physical processes controlling the flare productivity of ARs (one of the main goals of this study). To feasibly achieve this goal, we randomly selected five ARs from each group (circled in red in Figure \ref{fig:fi_area} and listed in Table \ref{table:ar_avg_parameters}). The sunspot areas of ARs are one of the basic parameters that characterize ARs. Therefore, we selected the ARs of interest on the basis of FI and the maximum sunspot area attained. It is important to note that though the boundaries are used to divide ARs into different groups, the selected ARs are distributed across a wide range of size and FI distributions.

For the selected ARs, we derived several relevant magnetic feature parameters. As the data quality of HMI vector magnetograms degrades considerably beyond 60$^{\circ}$ (see \citealt{Hoeksema_etal_2014}), we restricted the derivation of parameters within 60$^{\circ}$ of longitude from the central meridian. The derivation of the parameters is explained below. 

{\bf \textit{Total unsigned magnetic flux (USF):} }The total magnetic flux is a good quantitative measure of the total magnetic content of an AR. We use the absolute z-component of the magnetic field ($|B_{z}|$) to calculate the total flux, i.e., $\Phi$$_{tot}$$ = $$\sum_{i=1}^{n}{|{B_{z_{i}}}|}$$\Delta A$, where $\Delta A$ is the area of each pixel and $n$ is the total number of pixels.

{\bf \textit{Total unsigned vertical current ($I_{c}$):}}
Total unsigned vertical current ($I_{c}$) is one of the physical parameters that provide information about the extent of nonpotentiality of ARs (e.g., \citealt{Leka_and_Barnes_2003a}). Using vector magnetogram data that provide $B_{x}$, $B_{y}$ and $B_{z}$, we calculated the total unsigned vertical current at the photosphere as $I_{c}$=
$\sum_{i=1}^{n} |{J_{z_{i}}}|\Delta A$
=$\sum_{i=1}^{n} |\frac{1}{\mu_{o}}(\frac{\partial B_{y}}{\partial x}-\frac{\partial B_{x}}{\partial y})|\Delta A$.

{\bf \textit{Neutralized current:}} In relatively force-free-field plasma, the current through a flux tube (direct current) should be equal and opposite to the current flowing in the interface between the flux tube and the surrounding plasma (return current; \citealt{Parker_1996}). Thus, the net current at the photosphere, where flux tubes emerge and are surrounded by plasma, should be zero. Net current deviates from neutrality for ARs with local magnetic shear along PILs and such ARs have a higher rate of flare and coronal mass ejections (CMEs; e.g., \citealt{Georgoulis_etal_2012};  \citealt{Kontogiannis_etal_2017}; \citealt{Liu_etal_2017}). We used the ratio of dominant current to nondominant current ($|$DC/RC$|$) as a measure of the degree of current neutralization in each polarity (e.g. \citealt{Vemareddy_2015_flx_emer}; \citealt{Dhakal_etal_2020}). The ratio would be one in a current neutralized AR.

{\bf \textit{Current helicity ($H_{c}$):}}  Magnetic helicity is a topological measure of structural complexity of magnetic fields (\citealt{Berger_and_Field_1984}). It represents the linkages of magnetic fields i.e. the internal twist and external linking and knotting of flux tubes. The twist of magnetic fields of ARs provides insight into the nonpotentiality of ARs. Magnetic helicity is conserved in ideal MHD and roughly conserved in the presence of finite resistivity (e.g., \citealt{Berger_2005}). In the presence of finite resistivity, the temporal variation in the total magnetic helicity can be approximated by the total current helicity ($H_{c}$; e.g., \citealt{Leka_and_Barnes_2003a}). $H_{c}$ describes the linkages of electric currents and its rapid variation is linked with the origin of flares (e.g., \citealt{Bao_etal_1999}). We calculated the total current helicity (unsigned) over the ARs as $H_{c}$=$\sum_{i=1}^{n} |$$ B_{z_{i}}J_{z_{i}}|$.

{\bf \textit{Strong gradient PIL (SgPIL):}} PIL is an imaginary line that separates the opposite magnetic polarities. Shearing motion and flux cancellation along PILs can cause solar eruptions (e.g., \citealt{Green_etal_2011}). In this study, we analyzed the SgPIL of ARs using $B_{los}$. First, we determined the bitmaps of the positive or negative fluxes. These maps are dilated and then eroded with kernels of 15 $\times$ 15 and 5 $\times$ 5 pixels, respectively. Next, we shifted the bitmaps of magnetic fluxes by 10 pixels in both the $x$ and $y$ directions to determine the regions where the maps of opposite magnetic polarities overlapped with each other and where the gradient of the magnetic field was greater than 150 G Mm$^{-1}$. We discarded the isolated PIL where the length of the overlapped region is smaller than 10 pixels.

{\bf \textit{$R$ value:}} The $R$ value is the total amount of unsigned flux around PIL, and it provides the compactness of magnetic fluxes around PIL (\citealt{Schrijver_2007}). We calculated the $R$ value (using $B_{los}$) within $\sim$ 15 Mm of SgPIL. To calculate the $R$ value, we multiply the absolute value of the magnetogram with a weighting map. The weighting map is determined by taking the bitmap of SgPIL and convolving it with an area-normalized Gaussian with a FWHM of 15 Mm (similar to \citealt{Schrijver_2007}).

\begin{table}
\caption{Average Value of Magnetic Field Parameters of Selected ARs}  \label{table:ar_avg_parameters}
\begin {center}
\begin{tabular}{ |c|c|c|c|c|c|c|c|c|c|}
\hline
\multicolumn{10}{|c|}{{ Average Values of Magnetic Field Parameters}} \\
\hline
 AR		&	FI		&  SS 	&   FI 		&	USF		&	$I_{c}$		&	$H_{c}$ 					&	$|$DC/RC$|$		&	PIL 				&	$R$ value\\
		&	   		&    (MSH)&   (new)		&	(10$^{22}$ Mx)			&	(10$^{13}$ A)	&	(G$^{2}$ m$^{-1}$)			&			&	(Mm)				&	\\ 
 \hline
11429	& 	 103.7 	&   1270 	&     135.1		&      	5.1		&     5.8	  		&    4142.6	 				&     	 1.8  	&		103.6		&    5.0\\
11748	& 	 107.4  	&  310 	&      9.8		&      	2.0	  	&    3.2	   		&   1450.9	 					&     	 1.5  	&	  	17.0	    		&  4.0\\
12297	& 	 79.2    	&  420	&     100.1	 	&      	4.6		&     5.7	  		&    3648.7	  				&    	 1.5  	&		78.0	  		&    5.0\\
11967	& 	  76.6   	&  1580    	&     75.1		&      	10.6		&     11.5	   		&  9843.7	 					&    	 1.2 	&	  	146.0	   	&   5.3\\
11302	& 	 83.8    	&  1300    	&     92.6	 	&      	5.8	 	&     7.2	   		&   4271.7	  					&    	 1.2	&	  	71.9	   		&   4.9\\
11515	& 	 76.1    	&  900 	&     99.2	 	&      	6.4	 	&     7.7	  		&    4749.6	 				&     	 1.1	&	  	94.4	  		&    4.9\\
11875	& 	 64.0    	&   790 	&     38.4		&      	4.6		&     6.1	   		&   3365.3	  					&    	 1.1  	&	 	55.9	   		&   4.7\\

11158	& 	 58.3   	&   620 	&     74.0	 	&      	3.2		&     5.1	  		&    2933.6	  				&    	 1.2  	&	 	56.6	    		&  4.8\\

11928	& 	 33.6    	&   460  	&     37.6	 	&      	3.8		&    5.1	   		&   2734.5	  					&   	 1.0  	&	 	64.4	   		&   4.7\\
12205	& 	 38.4    	&  410	&     27.7		&      	4.7		&    5.8	   		&   3297.3	 					&     	 1.2  	&		59.0	   		&   4.7\\
11654	& 	 11.7    	&  1100 	&     15.6		&      	7.2	 	&    7.8	  		&    4380.2	  				&    	 1.0  	&		73.1	   		&   4.8\\
12209	& 	14.2 	   	&  1100    	&      19.8		&      	11.7   	&    10.2	  		&    6406.8	  				&    	 1.1  	&		61.4	   		&   4.8\\
11785	& 	10.1 	   	&  720 	&      11.0		&      	2.8    	&     3.8	  		&    1672.1	  				&   	 1.0 	&	  	17.2	    		&  4.0\\
12277	& 	9.3 	   	&  510 	&      10.0		&      	8.1		&     8.0	    		&  4162.2	  					&   	 1.0	&	   	30.9	    		&  4.3\\

12085	& 	 10.0    	&  840 	&    9.2		&      	3.8		&    4.7	   		&   2424.0	  					&    	 1.0 	&	  	36.4	    		&  4.5\\
11745	& 	 9.7    	& 600   	&    0.8		&      	5.6	 	&    5.6	    		&  2610.4	 					&     	 1.1 	&	  	19.8	    		&  4.1\\
11363	& 	5.2 		&  620 	&     2.4		&      	4.7	 	&     5.5	    		&  2567.3	  					&    	 1.1 	&	  	30.1	   		&   4.3\\
11917	& 	 4.3    	&  420 	&    3.5		&      	5.0		&    5.4	   		&   2439.2	  					&    	 1.1	&	   	19.0	   		&   4.1\\
12108	& 	 2.7    	&  890 	&    1.8		&      	4.8		&    5.3	   		&   2696.9	  					&   	 1.0	&	   	33.5	    		&  4.4\\
11430	& 	 2.9   	&   200 	&    4.0		&      	1.7		&     2.8	   		&   1093.8	  					&    	 1.1 	&	  	13.6	    		&  4.0\\

\hline
\end{tabular}
\end{center}
\footnotesize{\textbf{Note.} The first column shows the NOAA AR number. The second column shows FI for the front-disk transit of ARs. The third column shows the (peak) sunspot area. The fourth column shows FI within 60$^{\circ}$ of the central meridian. The fifth column shows the average of unsigned magnetic flux (USF). The sixth column shows the average of vertical current ($I_{c}$). The seventh column shows the average of current helicity ($H_{c}$). The eighth column shows the average of $|$DC/RC$|$. The ninth column shows the average of the length of  strong-gradient polarity inversion line. The tenth column shows the average of $R$ value.}
\end{table}

\section{\textbf{On the Contrasting Evolution of Four Types of ARs}}
\label{AR_evolution}
\begin{figure}[!ht]
\centering
\includegraphics[width=.81\textwidth,clip=]{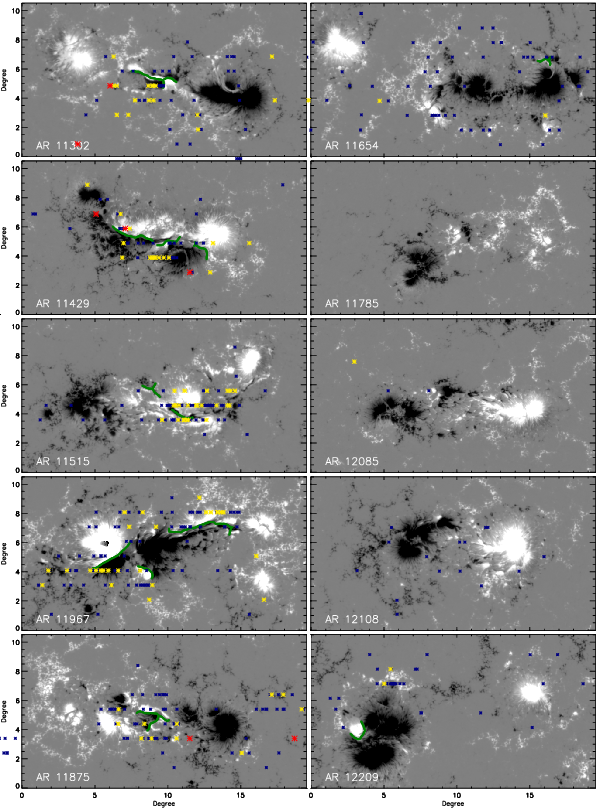}
\caption{Snapshot images of the 10 selected ARs with large sunspot areas. The positive (negative) magnetic fluxes are shown in white (black). The left panels show ARs with super flare productivity (SAARs). The right panels show ARs with moderate flare productivity (MAARs). The locations of C-, M-, and X-class flares (during the entire front-disk transit period) are shown in blue, yellow, and red asterisks, respectively. The green lines show the locations of the strong-gradient polarity inversion line.}
\label{fig:lar_ars}
\end{figure}

\begin{figure}[!ht]
\centering
\includegraphics[width=.81\textwidth,clip=]{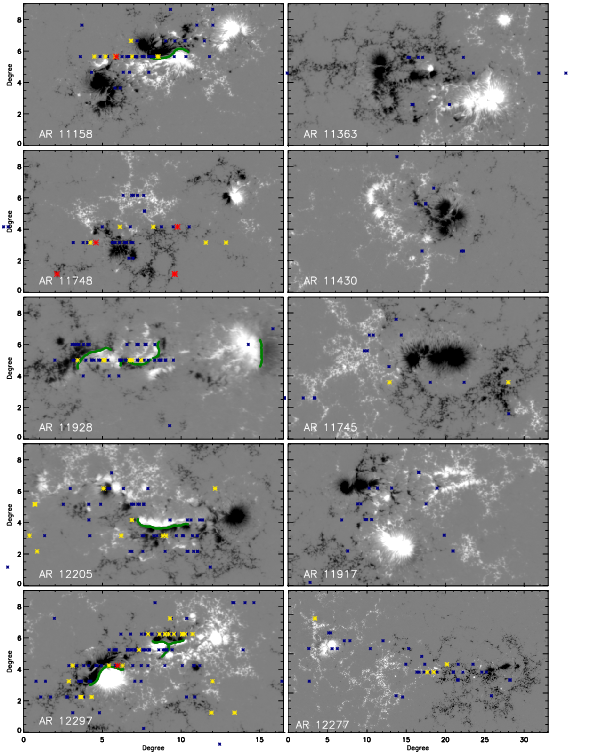}
\caption{Snapshots images of the 10 selected ARs with small sunspot areas. The positive (negative) magnetic fluxes are shown in white (black). The left panels show ARs with super flare productivity (SAARs). The right panels show ARs with moderate flare productivity (MAARs). The locations of C-, M-, and X-class flares (during the entire front-disk transit period) are shown in blue, yellow, and red asterisks. The green lines show the location of the strong-gradient polarity inversion line. }
\label{fig:sar_ars}
\end{figure}

Figures \ref{fig:lar_ars} and \ref{fig:sar_ars} show the ARs with large and small sunspot sizes, respectively. The left panels of these images show SAARs, and the right panels show MAARs. Irrespective of their sizes, SAARs have complex magnetic configurations, where opposite magnetic fluxes are close to each other to form a long SgPIL in the middle of ARs (see the left panels of Figures \ref{fig:lar_ars} and \ref{fig:sar_ars}). Comparatively, MAARs have relatively (overall) simple magnetic configurations, where the groups of opposite magnetic poles are far from each other and lack long SgPIL (see the right panels of Figure \ref{fig:lar_ars} and \ref{fig:sar_ars}). Another striking observational difference between them is the location of flares. SAARs have multiple intense (M- and X-class) flares originating around the SgPIL (see Figures \ref{fig:lar_ars} and \ref{fig:sar_ars}). No such obvious pattern is observed for the location of flares in MAARs. It is noteworthy that Figures \ref{fig:lar_ars} and \ref{fig:sar_ars} are reflecting the magnetic configurations of ARs only at a particular time, whereas flares occurred throughout the front-disk passage of ARs. Moreover, the flare locations are determined using coronal data (e.g., SXR) and thus may not accurately superimpose on the phototospheric images. Nevertheless, it is clear that most of the intense flares occurred around the SgPIL in SAARs.

Most of the selected ARs appeared on the front disk due to solar rotation. Only three ARs (AR 11158, 11928, and 12085) were newly emerged on the front disk. New emergence of magnetic fluxes has a higher tendency to occur near or within existing ARs, known as the nesting tendency of flux emergence (see \cite{van_Driel_and_Green_2015} and references therein). Magnetic flux emergence of significant magnitude was observed, near the existing magnetic flux, in almost every AR (except for AR 11748, 11745, 12277, and 12209). However, new emergence was not always associated with intense flares or an increase in flare productivity. The following examples of four ARs, one from each group (discussed in Section \ref{instru}), show in detail when the new emergence changes the magnetic configuration and consequently flare activity in ARs. Two of the four selected ARs, AR 11430 and 12108, that are associated with only C-class flares represent MAARs. The other two ARs, AR 11515 and 11928, that are associated with  multiple M-class flares represent SAARs.

\subsection{Evolution of AR 11430 of a Small Sunspot Area and Low FI}
\begin{figure}[!ht]
\centering
\includegraphics[width=1\textwidth,clip=]{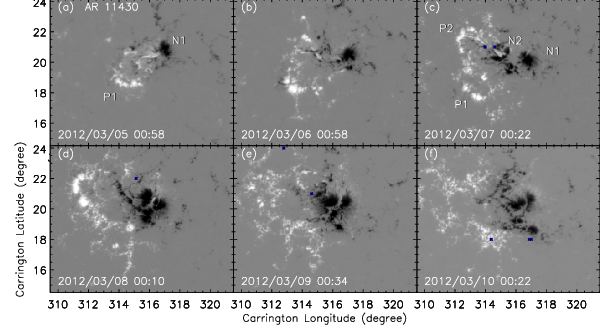}
\caption{Evolution of AR 11430 as observed in the HMI line-of-sight magnetogram in the period of 6 days. Positive (negative) flux is shown in white (black). The locations of C-class flares are shown in blue asterisks. An animation showing the evolution of the AR, between 2012 March 04, 19:58 and March 12, 16:58 UT, is available. The duration of the video is 30 s.}
\label{fig:evo_11430_new}
\end{figure}

AR 11430, from group I (discussed in section \ref{instru}), had a small sunspot size (200 MSH at the peak) and small FI (2.9). It was first observed on 2012 March 5, with a simple $\beta$ magnetic configuration. Table \ref{table:ar_11430} shows the key parameters summarizing the magnetic evolution and flare history of AR 11430. The total magnetic flux content (unsigned) is shown only when the AR was located within 60$^{\circ}$ of longitude from the central meridian. On March 6, a new magnetic bipole (P2-N2) emerged from the side of the existing magnetic bipole (P1-N1; see Figure \ref{fig:evo_11430_new}, see also the accompanying movie). The orientation of the newly emerging magnetic poles was consistent with Hale's law. The separation motion during the emergence led the emerging negative polarity to move and merge with the existing negative poles. Few episodes of small flux emergence were observed between March 8 and 10. However, the emerging poles were not strong enough to form new sunspots. Therefore, the total sunspot area of the AR decreased after March 8 (see Table \ref{table:ar_11430}). The opposite magnetic fluxes diffused and came close to each other during the decay phase (see Figure \ref{fig:evo_11430_new}(e) and (f)).

\begin{table}[h]
\caption{Changes of Key Parameters Showing the Evolution of AR~11430}
\label{table:ar_11430}
\begin {center}
\begin{tabular}{ |c|c|c|c|c|c|c|}
 \hline
 \multicolumn{7}{|c|}{{\bf NOAA AR 11430}} \\
 \hline
 Date & Area (MSH) & USF (10$^{22}$ Mx)   & X & M & C & Sunspot Type\\
 \hline
 2012/03/05& 20&    1.6 &	0&	0&	0&	$\beta$\\
 2012/03/06& 90&    1.5 &	0&	0&	2&	N/A\\
 2012/03/07& 110&   1.6    &	0&	0&	1&	$\beta$\\
 2012/03/08& 200&   1.9    &	0&	0&	2&	$\beta$\\
 2012/03/09& 180&   1.8    &	0&	0&	1&	$\beta$\\
 2012/03/10& 120&   2.0    &	0&	0&	1&	$\beta$\\
 2012/03/11& 100&   2.2    &	0&	0&	0&	$\beta$\\
 2012/03/12& 100&   2.5    &	0&	0&	1&	$\beta$\\
 2012/03/13& 30&    -    &	0&	0&	0&	$\alpha$\\
 2012/03/14& 30&     -   &	0&	0&	0&	$\alpha$\\
\hline
\end{tabular}
\end {center}
\end{table}

Throughout the observational period, the overall magnetic configuration of AR 11430 was a simple $\beta$ configuration, where opposite magnetic polarities were well separated from each other. The new flux emergence occurred in such a way that there was no interaction between opposite magnetic fluxes of emerging and existing magnetic polarities. There were only C-class flares from this AR (see Table \ref{table:ar_11430}) and flare productivity was similar during both the emergence and the decay phase.

\subsection{Evolution of AR 12108 of a Large Sunspot Area and Low FI}
\begin{figure}[!ht]
\centering
\includegraphics[width=1\textwidth,clip=]{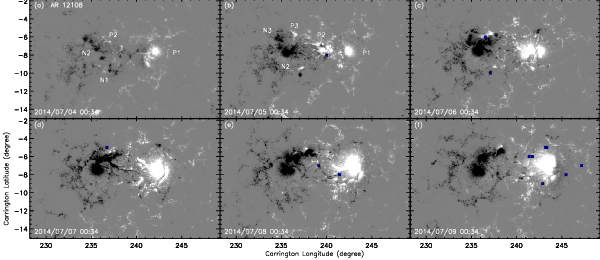}
\caption{Evolution of AR 12108 as observed in the HMI line-of-sight magnetogram in the period of six days. Positive (negative) flux is shown in white (black). The locations of C-class flares are shown in blue asterisks. An animation showing the evolution of the ARs, between 2014 July 02, 11:58 and July 10, 00:58 UT, is available. The duration of the video is 29 s.}
\label{fig:evo_12108_new}
\end{figure}

AR 12108, from group II (discussed in section \ref{instru}), had a large sunspot area (890 MSH at the peak) and a relatively small FI (2.7). It had a simple $\beta$ magnetic configuration as it appeared on the eastern limb on 2014 July 1. Table  \ref{table:ar_12108} shows the key parameters summarizing the magnetic evolution and flare history of AR 12108. On July 2, a new magnetic bipole (P2-N2) emerged in between the existing magnetic bipole (P1-N1; see Figure \ref{fig:evo_12108_new}(b) and the accompanying movie). The orientation of the newly emerging magnetic poles was consistent with Hale's law. As the emergence was in the middle of the existing magnetic bipole, the separation of emerging fluxes led the emerging positive (negative) fluxes to move and merge with the existing positive (negative) poles. There was another emergence of a small magnetic bipole (P3-N3) toward the north of P2-N2 on July 4. Due to the location of the emerging P3-N3, their separation motion led P3 to move toward the N2. The interaction of P3 and N2 resulted in a complex magnetic configuration ($\beta/\gamma/\delta$) of the AR for about three days. However, due to the small size of P3, the interaction between the opposite magnetic fluxes of nonconjugate poles (P3 and N2) lasted only for a brief period of time. The interacting region was also small. Only, C-class flares were observed around this region (see panels (c) and (d) of Figure \ref{fig:evo_12108_new}).

\begin{table}
\caption{Changes of key parameters showing the evolution of AR 12108}
\label{table:ar_12108}
\begin {center}
\begin{tabular}{ |c|c|c|c|c|c|c|}
 \hline
 \multicolumn{7}{|c|}{{\bf NOAA AR 12108}} \\
 \hline
 Date & Area (MSH)  & USF (10$^{22}$ Mx)  &   X   & M     & C     & Sunspot Type\\
 \hline
 2014/07/01&  N/A   & -	&   0   &	0   &	1   &	N/A\\
 2014/07/02& 30     & -	&	0   &	0   &	0   &	$\beta$\\
 2014/07/03& 90     & 4.9	&	0   &	0   &	0   &	$\beta$\\
 2014/07/04& 90     & 4.1	&	0   &	0   &	1   &	$\beta$\\
 2014/07/05& 120    & 3.9	&	0   &	0   &	2   &	$\beta/\gamma$\\
 2014/07/06& 350    & 4.0	&	0   &	0   &	1   &	$\beta/\gamma$\\
 2014/07/07& 620    & 4.4	&	0   &	0   &	2   &	$\beta/\gamma/\delta$\\
 2014/07/08& 720    & 4.9	&	0   &	0   &	0   &	$\beta/\gamma/\delta$\\
 2014/07/09& 890    & 5.4	&	0   &	0   &	0   &	$\beta/\gamma/\delta$\\
 2014/07/10& 830    & 6.1	&	0   &	0   &	1   &	$\beta/\gamma/\delta$\\
 2014/07/11& 690    & 7.4	&	0   &	0   &	0   &	$\beta/\gamma$\\
 2014/07/12& 560    & -	&	0   &	0   &	3   &	$\beta/\gamma$\\
 2014/07/13& 600    & -	&	0   &	0   &	3   &	$\beta/\gamma$\\
 2014/07/14& 200    & -	&	0   &	0   &	1   &	$\beta/\gamma$\\
 2014/07/15& N/A    & -	&	0   &	0   &	1   &	N/A\\
 \hline
\end{tabular}
\end {center}
\end{table}

During most of the observational period, the negative poles were well separated from the positive pole in AR 12108 (see Figure \ref{fig:evo_12108_new}). The new flux emergence had occurred in such a way that there was no long-lasting interaction between opposite magnetic fluxes of emerging and existing magnetic polarities. There were only C-class flares from this AR (see Table \ref{table:ar_12108}), and the flare productivity was similar both during the emergence and the decay phase.

\subsection{Evolution of AR 11515 of a Large Sunspot Area and High FI}
\begin{figure}[!ht]
\centering
\includegraphics[width=1\textwidth,clip=]{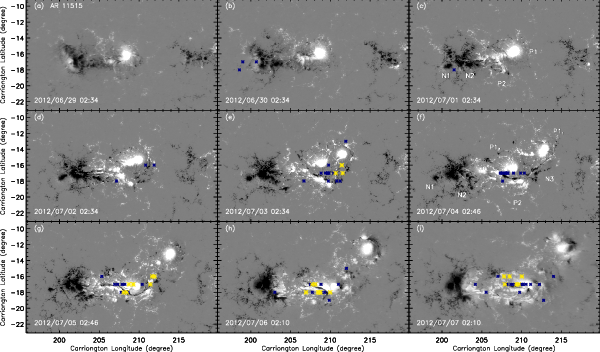}
\caption{Evolution of AR 11515 as observed in the HMI line-of-sight magnetogram in the period of nine days. Positive (negative) flux is shown in white (black). The locations of C-class and M-class flares are shown in blue and yellow asterisks, respectively. An animation showing the evolution of the AR, between 2012 June 28, 21:34 and July 07, 03:10 UT, is available. The duration of the video is 33 s.}
\label{fig:evo_11515_new}
\end{figure}

AR 11515, from group III (discussed in section \ref{instru}), had a large sunspot area (900 MSH at the peak) and high FI (76.1). Table \ref{table:ar_11515} shows the key parameters summarizing the magnetic evolution and flare history of AR 11515. It appeared on the eastern limb with a $\beta$ magnetic configuration on 2012 June 28.  Several episodic flux emergences occurred in the middle of existing bipole (P1-N1) throughout the observational period. P1-N1 is a simplified representation of the magnetic feature resulting from possible complex evolution on the back side of the Sun. During the separation of emerging fluxes, the negative (positive) fluxes coalesced to form N2 (P2; see Figure \ref{fig:evo_11515_new} and the accompanying movie). As the negative (positive) fluxes of subsequent episodic emergence, in the middle of existing poles, merged with N2 (P2), they are not labeled differently in Figure \ref{fig:evo_11515_new}. The episodic emergence, in the middle of existing poles, is similar to the emergence observed in AR 11430 and AR 12108. Although episodic flux emergence was observed since June 28, there were only a few C-class flares from this region (see Figure \ref{fig:evo_11515_new}). However, the subsequent emergence was much more complex.

The existing positive pole (P1) has been moving in the northwest direction since the beginning of the observation. On July 1, around 14:00 UT, a new flux emergence occurred near P1. With the continuation of the flux emergence, P1 was divided into two parts (P1$_{1}$ and P1$_{2}$). The emerging positive flux seemed to merge with P1$_{1}$ and the emerging negative flux moved toward P2. P1$_{2}$ also moved toward P2. All the emerging magnetic pairs followed Hale's law of polarity; with the leading (following) pole with positive (negative) magnetic flux. Three M-class flares occurred during the emergence period near P1 and N3 (see Figure \ref{fig:evo_11515_new} (e)). Due to the convergence of P2, N3, and P1$_{2}$, a long PIL started to form in the middle of the AR (see panel (e) of Figure \ref{fig:evo_11515_new}). 19 M-class flares occurred near this subregion (see panels (g)-(i) of Figure \ref{fig:evo_11515_new}).

\begin{table}
\caption{Changes of Key Parameters Showing the Evolution of AR 11515.}
\label{table:ar_11515}
\begin {center}
\begin{tabular}{ |c|c|c|c|c|c|c|}
 \hline
 \multicolumn{7}{|c|}{{\bf NOAA AR 11515}} \\
 \hline
 Date & Area (MSH) & USF (10$^{22}$ Mx)   &X & M & C & Sunspot Type\\
 \hline
 2012/06/28&  200   & 6.8  &		0&	0&	1&	$\beta$\\
 2012/06/29& 180    & 6.7  &		0&	0&	3&	$\beta/\gamma$\\
 2012/06/30& 310    & 5.9  &		0&	0&	1&	$\beta$\\
 2012/07/01& 380    & 5.2  &		0&	0&	3&	$\beta/\gamma$\\
 2012/07/02& 850    & 4.8  &		0&	3&	11&	$\beta/\gamma$\\
 2012/07/03& 620    & 4.9  &		0&	0&	13&	$\beta/\gamma$\\
 2012/07/04& 570    & 5.7  &		0&	5&	10&	$\beta/\gamma/\delta$\\
 2012/07/05& 640    & 6.8  &		0&	8&	11&	$\beta/\gamma/\delta$\\
 2012/07/06& 670    & 8.3  &		0&	2&	6&	$\beta/\gamma/\delta$\\
 2012/07/07& 900    & 10.4  &		0&	2&	6&	$\beta/\gamma/\delta$\\
 2012/07/08& 780    & -  &			0&	4&	8&	$\beta/\gamma$\\
 2012/07/09& 550    & -  &			0&	0&	3&	$\beta/\gamma$\\
 2012/07/10& 320    & -  &			0&	0&	1&	$\beta/\gamma$\\
 \hline
\end{tabular}
\end {center}
\end{table}

AR 11515 had a simple $\beta$ configuration in the earlier evolutionary period. Later on, the magnetic configuration of the AR became very complex with the new emergence and interaction between nonconjugate opposite fluxes. The negative fluxes were moving toward the left and the positive fluxes were moving toward the right. This indicated a continuous shearing motion between opposite magnetic fluxes. Along with the shearing motion, a continuous flux cancellation was observed in this subregion (see panels (f)-(i) of Figure \ref{fig:evo_11515_new}). Many of the M-class flares originated around this subregion (the locations of flares are shown in yellow asterisks in Figure \ref{fig:evo_11515_new}).

\subsection{Evolution of AR 11928 of a Small Sunspot Area and High FI}
\begin{table}
\caption{Changes of Key Parameters Showing the Evolution of AR 11928.}
\label{table:ar_11928}
\begin {center}
\begin{tabular}{ |c|c|c|c|c|c|c|}
 \hline
 \multicolumn{7}{|c|}{{\bf NOAA AR 11928}} \\
 \hline
 Date & Area (MSH) & USF (10$^{22}$ Mx)  & X & M & C & Sunspot Type\\
 \hline
 2013/12/18&  130   & 2.2  &		0&	0&	1&	$\beta/\gamma$\\
 2013/12/19& 240    & 2.7  &		0&	0&	3&	$\beta/\gamma$\\
 2013/12/20& 360    & 3.2  &		0&	0&	5&	$\beta/\gamma$\\
 2013/12/21& 400    & 4.1  &		0&	0&	9&	$\beta/\gamma$\\
 2013/12/22& 460    & 5.2  &		0&	5&	11&	$\beta/\gamma$\\
 2013/12/23& 380    & -  	&		0&	1&	11&	$\beta/\gamma$\\
 2013/12/24& 330    &  - 	&		0&	0&	2&	$\beta/\gamma$\\
 2013/12/25& 130    &  - 	&		0&	0&	5&	$\beta/\gamma$\\ 
\hline
\end{tabular}
\end {center}
\end{table}

\begin{figure}[!ht]
\centering
\includegraphics[width=1\textwidth,clip=]{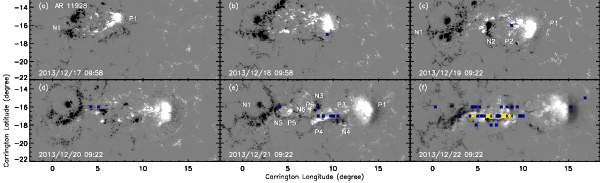}
\caption{Evolution of AR 11928 as observed in the HMI line-of-sight magnetogram in the period of six days. Positive (negative) flux is shown in white (black). The locations of C-class and M-class flares are shown in blue and yellow asterisks, respectively. An animation showing the evolution of the AR, between 2013 December 15, 23:58 and December 22, 20:22 UT, is available. The duration of the video is 27 s.}
\label{fig:evo_11928_new}
\end{figure}

AR 11928, from group IV (discussed in section \ref{instru}), had a small sunspot area (460 MSH at the peak) and high FI (33.6). The evolutionary summary of the AR is in Table \ref{table:ar_11928}. It was one of the ARs (from selected) that newly emerged on the front disk of the Sun. In an area of diffused negative polarity, probably from another but fully decayed AR, a new magnetic bipole (P1-N1) started to emerge on 2013 December 16, at $\sim$ 04:24 UT. The emerging fluxes coalesced into three different groups, due to which the AR acquired a multipolar configuration (see Table \ref{table:ar_11928} and panel (a) of Figure \ref{fig:evo_11928_new}). For simplicity, the group of negative polarity is labeled as N1. P1 and N1 were far from each other.  There were many subsequent magnetic emergences after December 20; P3-N3 emerged near P1, P4-N4 emerged near N2, P5-N5 emerged near P1, and P6-N6 emerged in the middle of the AR (see panel (e) of Figure \ref{fig:evo_11928_new} and accompanying movie). Except for P4-N4, all the other emerging magnetic pairs followed Hale's law of polarity; with the leading (following) pole having positive (negative) magnetic flux. In the middle of the AR, the interaction among nonconjugate pairs (P4, P5, P6, and N5) formed a long PIL. Although the size of each emerging bipole was small, long PIL indicated the interaction between opposite magnetic fluxes over a large area. Six M-class flares originated around this location (see panel (f) of Figure \ref{fig:evo_11928_new}).

In the above, we compared the evolution of two SAARs and two MAARs. In SAARs, the interaction between opposite magnetic poles of nonconjugate pairs results in the formation of PILs. Such interaction between nonconjugate pair were observed in almost every SAAR, except for AR 11748 and AR 11302.  Most of the intense flares from AR 11748 occurred when it was near the eastern limb. It had the magnetic configuration of decaying AR near the solar disk. In AR 11302, a positive flux region was sandwiched between the negative fluxes in the middle of ARs  (see Figure\ref{fig:lar_ars}). This region was formed near the eastern limb, and it was not clear that the opposite fluxes were from the nonconjugate pair. Also, the selection of the sample ARs in each of the four types is rather random. Thus, they are representative of their types.

\section{Evolution of the Magnetic Feature Parameters of Four Types of ARs}
\label{flare_driver}
The above sections discussed the contrasting evolution of SAARs and MAARs based on visual inspection of the magnetograms. To quantify the contrasting evolution, we calculated and analyzed six magnetic feature parameters (discussed in Section \ref{instru}) within 60$^{\circ}$ of longitude from the central meridian. The time profile of each parameter is discussed below.

\begin{figure}[!ht]
\centering
\includegraphics[width=.63\textwidth,clip=]{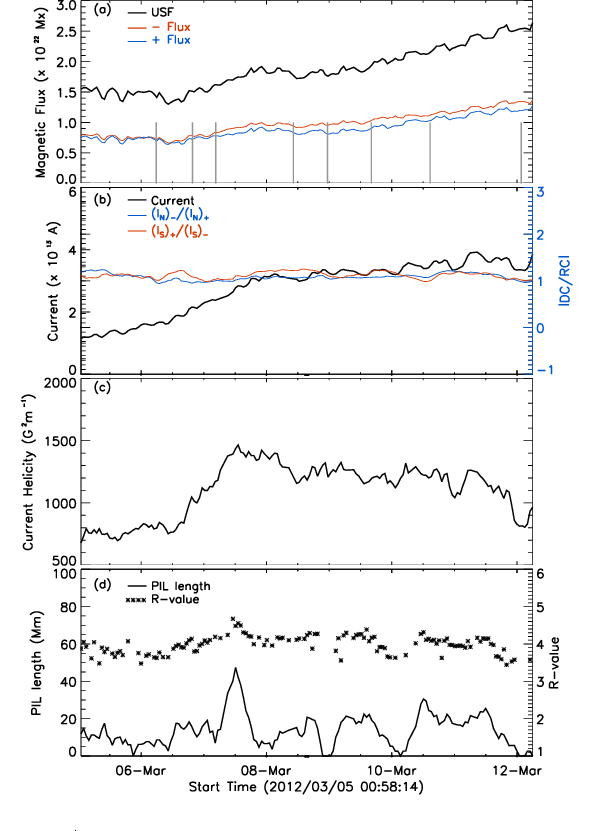}
\caption{Evolution of the magnetic feature parameters in AR 11430. (a) Time-profiles of total unsigned, positive, and negative magnetic flux of the AR in black, blue, and red, respectively. The vertical gray (green) lines indicate the times of C-class (M-class) flares from the AR. (b) Profiles of the total unsigned current in black and $|$DC/RC$|$ in positive (negative) polarity in blue (red). (c) Profile of current helicity in black. (d) Length of strong-gradient polarity inversion line (SgPIL) and $R$ value in solid black line and black asterisks, respectively.}
\label{fig:11430_non_pot}
\end{figure}

\subsection{Total Unsigned Magnetic Flux (USF)} 
Panel (a) of Figures \ref{fig:11430_non_pot}-\ref{fig:11928_non_pot} shows the temporal variation in USF in black and positive (negative) magnetic flux in blue (in red) in ARs 11430, 12108, 11515, and 11928, respectively. Each AR had multiple episodes of magnetic flux emergence (as discussed in Section \ref{AR_evolution}). There is no obvious relation between the rate of change of the magnetic flux and flare productivity from these ARs. This suggests that flare productivity does not change only with the variation in the total flux content. This result is consistent with the discussion in Section \ref{AR_evolution} in the cases where the emerging polarities simply merge with the existing poles of like polarity.

\subsection{Total Unsigned Current ($I_{c}$)} 
Panel (b) of Figures \ref{fig:11430_non_pot}-\ref{fig:11928_non_pot} shows the time profile of $I_{c}$ (in black) of ARs 11430, 12108, 11515, and 11928, respectively. The total current in AR 11430 increased at a rate of 7.9$\times$ 10$^{7}$ A s$^{-1}$ until it reached the maximum sunspot size on 2012 March 8, afterward, it did not vary that much. The total current in AR 12108 increased gradually throughout the observational period at a rate of 7.8$\times$ 10$^{7}$ A s$^{-1}$. The current in AR 11515 was also increasing throughout the observation period at a rate of 1.0 $\times$ 10$^{8}$ A s$^{-1}$. Similarly, the total current was increasing in AR 11928 at a rate of 8.2$\times$ 10$^{7}$ A s$^{-1}$ throughout the observational period. The rate of increase of the current was higher in AR 11515 compared to other three ARs. The increase rate of current was comparable for ARs 11430, 12108, and 11928. Also, the total current increased in all four ARs. However, flare productivity did not increase correspondingly. 

\subsection{Degree of Current Neutralization}
\begin{figure}
\centering
\includegraphics[width=.63\textwidth,clip=]{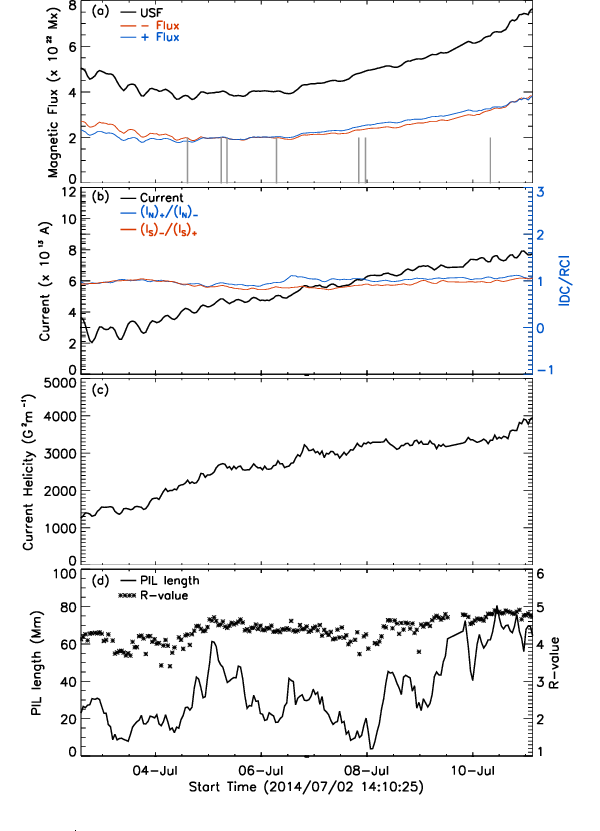}
\caption{Evolution of the magnetic feature parameters of AR 12108. The caption is the same as in Figure \ref{fig:11430_non_pot}.}
\label{fig:12108_non_pot}
\end{figure}

\begin{figure}
\centering
\includegraphics[width=.63\textwidth,clip=]{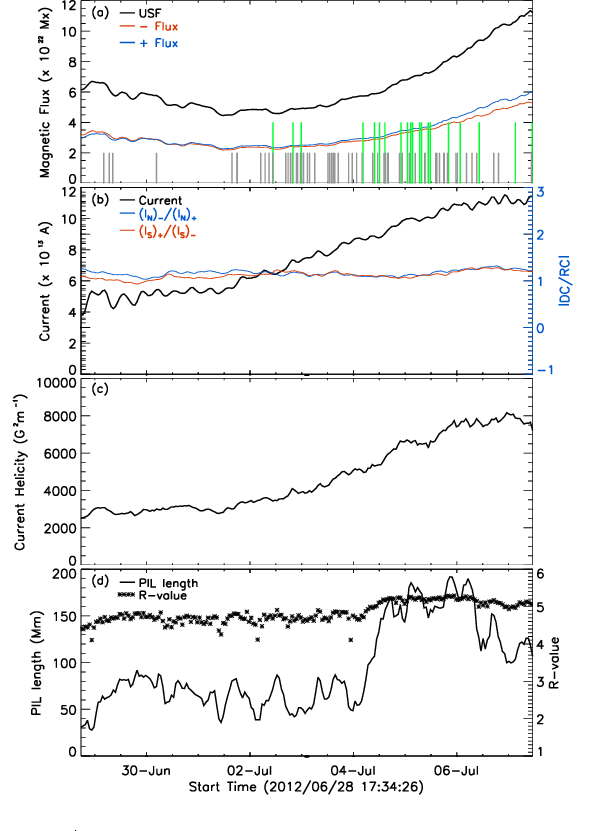}
\caption{Evolution of the magnetic field parameters in AR 11515. This is similar to Figure \ref{fig:11430_non_pot}.}
\label{fig:11515_non_pot}
\end{figure}

Panel (b) of Figures \ref{fig:11430_non_pot}-\ref{fig:11928_non_pot} also shows the evolution of $|DC/RC|$ in positive (negative) polarity of  ARs 11430, 12108, 11515, and 11928, respectively in blue (red). The ratio was undulating for AR 11430; however, flare productivity was almost the same during the rise and fall of the ratio. For AR 12108, the ratio was close to unity throughout the observational period. The ratio was also undulating for AR 11515 during the observational period. In AR 11928, intense flares were observed when the ratio was decreasing and approaching unity. There was no obvious pattern between flare productivity and the rate of change of the ratio.

\subsection{Current Helicity ($H_{c}$)} 
\begin{figure}[!ht]
\centering
\includegraphics[width=.62\textwidth,clip=]{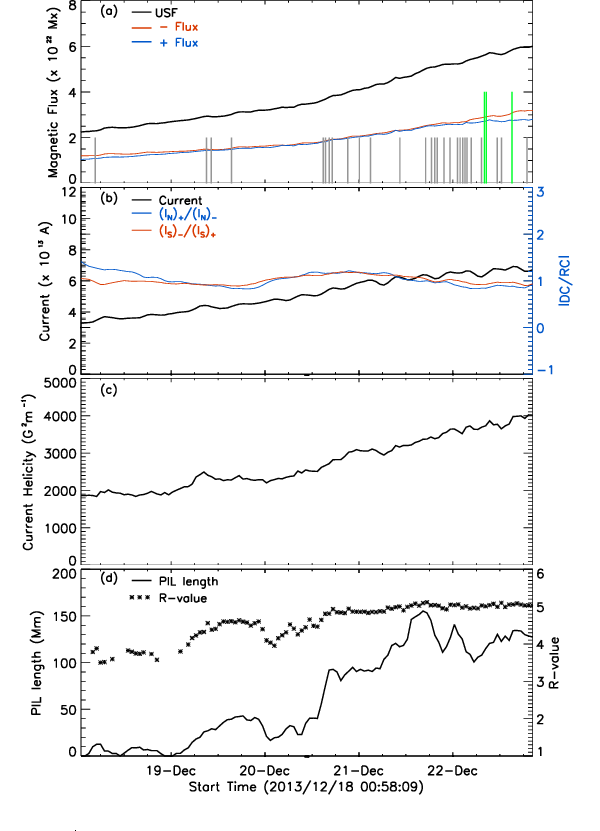}
\caption{Evolution of the magnetic field parameters in AR 11928. This is similar to Figure \ref{fig:11430_non_pot}.}
\label{fig:11928_non_pot}
\end{figure}

Panel (c) of Figures \ref{fig:11430_non_pot}-\ref{fig:11928_non_pot} shows the time profile of $H_{c}$ of  ARs 11430, 12108, 11515, and 11928, respectively. Although the $H_{c}$ increased at a rate of  2.7$\times$10$^{-3}$ G$^{2}$ m$^{-1}$ s$^{-1}$ until 2012 March 8 in AR 11430, afterward it decreased slowly. $H_{c}$  in ARs 12108, 11515, and 11928 increased throughout the observational period. $H_{c}$ in AR 11515 was much higher ($\sim$8000 G$^{2}$ m$^{-1}$ at the end of the observational period) than the other three ARs. However, it was comparable ($\sim$4000 G$^{2}$ m$^{-1}$ near the end of the observational period) for ARs 12108 and 11928. This suggests that the increase in $H_{c}$  or the magnitude of $H_{c}$  may not obviously reflect the frequency or intensity of flares.

\subsection{SgPIL and $R$ value} 
Panel (d) of Figures \ref{fig:11430_non_pot}-\ref{fig:11928_non_pot} shows the variation in the length of SgPIL of  ARs 11430, 12108, 11515, and 11928, respectively in black solid lines. AR 11430 had a very small SgPIL ($<$ 50 Mm) throughout the observational period. For most of the observational period, the length of SgPIL was $<$ 40 Mm. Although there was a jump in SgPIL ($>$ 40 Mm) for less than a day (between 2012 March 7 and 8), there were no intense flares from the AR during this period. AR 12108 also had a small SgPIL (maximum of $\sim$70 Mm). There was an increase in the SgPIL (up to $\sim$60 Mm) between 2012 March 4 and 6. However, there were no intense flares from the AR. ARs 11515 and 11928 had small SgPIL during the initial evolutionary period. However, the length of their SgPIL increased a lot in the later evolution. While there were a few intense flares in AR 11515 prior to the jump in the length of SgPIL, the flare rate increased significantly during and after the jump. In AR 11928 the number of C-class flares increased with the jump in the length of SgPIL. However, there were no intense flares during the rise in the length of SgPIL. There were intense flares in AR 11928 only after the SgPIL reached the maximum value. As the $R$ value represents the compactness of opposite magnetic fluxes near the SgPIL, it follows a similar trend as that of SgPIL. Unlike MAARs, SAARs had longer SgPIL and the flare activity of these ARs seems to increase with the length of the SgPIL/ $R$ value.

In short, we compared the variation in the magnetic feature parameters with the evolution of ARs and their flare productivity. The most striking result is that a considerable increase in PIL length/ $R$ value leads to an increase in flare productivity. For other parameters, there was no such obvious relation.

\subsection{Correlation of Flare Productivity with Magnetic Feature Parameters} 
In addition to analyzing the changes in flare productivity with the variation in the magnetic feature parameters, we wanted to investigate how, on average, the flare productivity changes with the magnitude of feature parameters.
For this reason, we calculated the average value of magnetic feature parameters during the observational period of each AR to compare with flare productivity.  
Table \ref{table:ar_avg_parameters} shows the time-averaged value of magnetic field parameters in each AR. As these parameters were calculated within 60$^{\circ}$ of the central meridian, we calculated FI (new) based on the flares observed within 60$^{\circ}$ of the central meridian and compared them with the parameters.

\begin{figure*}[!ht]
\centering
\includegraphics[width=.8\textwidth,clip=]{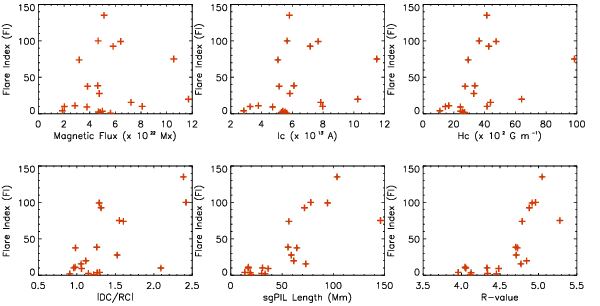}
\caption{Scatter plots between FI and the magnetic field parameters for the selected ARs. The top left, middle, and right panels show plots for the total unsigned magnetic flux, current density, and current helicity, respectively. The bottom left, middle, and right panels show plots for $|$DC/RC$|$, the length of SgPIL, and the $R$ value, respectively.}
\label{fig:avg_para_fi}
\end{figure*}

Figure \ref{fig:avg_para_fi} shows the scatter plots between the time-averaged magnetic feature parameters and FIs. Plots of SgPIL and the $R$ value are less scattered compared to the plots of the magnetic flux, total unsigned current, current helicity, and $|$DC/RC$|$. Also, the scatter plot of the SgPIL/ $R$ value shows the division of ARs into two groups. ARs with SgPIL $<$ 50 Mm (R-value $<$ 4.5) have relatively small FIs (see Figure \ref{fig:avg_para_fi}). We calculated the correlation coefficients between FI and magnetic feature parameters to understand their relation (see Table \ref{table:correlation_fi_parameters}).  Our selection of ARs was based on the contrasting types of FI sand sunspot areas (at peak), which is reflected by a correlation coefficient of  $\sim$0.5 between these two.
 Among all the magnetic feature parameters, the total unsigned flux content has the weakest correlation with FI (0.14). As the total flux contents of ARs reflect their sizes (e.g., \citealt{Sheeley_1966}), this result suggests that large ARs do not necessarily produce intense flares and vice versa. The correlation between the total vertical current and FI is 0.28, and the correlation between the current helicity and FI is 0.44. In each AR, episodes of new flux emergence were observed. An emerging flux tube injects current and helicity into the corona. The total current and current helicity of the ARs increased with the new flux emergence. However, as discussed above such an increase in the total current content and the current helicity does not lead to the increase in the flare activity. This is reflected by the weak correlation between these parameters and FI. The correlation between $|$DC/RC$|$ and FI is 0.64. Long SgPIL is the distinctive feature of SAARs, and therefore nonneutralized current has a moderate correlation. The $R$ value, which is closely related to SgPIL, is also strongly correlated with FI (with a correlation coefficient of 0.75). It is interesting to point out that the correlation between FI and the length of SgPIL is the highest (0.78). This suggests that ARs with long SgPIL are most likely to produce intense flares and be flare productive. This is similar to the finding of \cite{Schrijver_2007}, who found that the probability of intense flare increases with the length of SgPIL. They suggest that SgPIL is the characteristic of emerging magnetic structures. However, we argue that the photospheric evolution leading to the interaction between the opposite fluxes of nonconjugate pair plays a more important role than the flux emergence.

\vspace{15 mm}

 \begin{table}
 \caption{Correlation between FI and the Average of Magnetic Field Parameters}
\label{table:correlation_fi_parameters}
\begin {center}
 \centering
\begin{tabular}{ |c|c|c|c|c|c|c|c|}
 \hline
 \multicolumn{8}{|c|}{${\bf Correlations}$} \\
 \hline
 Correlation & Total Flux & $I_{c}$    &  $H_{c}	$	&	Sunspot Area$^{a}$   & $|$DC/RC$|$ & $R$ value & SgPIL\\
 \hline
 FI& 0.14 & 0.28 & 0.44 & 0.47 & 0.69 & 0.75 & 0.78\\
\hline
\end{tabular}
\end {center}
\footnotesize{\textbf{Note.} $^{a}$ Peak sunspot area within  60$^{\circ}$ of central meridian.}
\end{table}

\section{\textbf{Discussion and Conclusions}}
\label{DC}
 
In this study, we analyzed and compared the evolution of ten SAARs and ten MAARs. Here, we used FI and the sunspot size to determine and classify ARs into different activity groups. Although the boundaries were used to divide ARs into four groups for the initial selection convenience, the selected ARs had wide range of sizes and FIs. Also, for the quantitative analysis, all the ARs were considered together. Thus, we believe that the choice of boundaries has no or minimum impact on the results regarding the superactivity of ARs. One of the main objectives of this study was to identify the physical processes responsible for multiple intense flares from an AR. We found that the presence of long SgPILs in the middle of ARs was the most distinctive feature for the SAARs, whereas in MAARs opposite magnetic poles were relatively far from each other. The SgPILs are one of the important observational features identified in flare productive ARs for a long time (e.g., \citealt{Zirin_and_Wang_1993}; \citealt{Vemareddy_2019}). The shearing motion and flux cancellation of opposite magnetic fluxes along the PIL are well known to produce solar eruptions (e.g., \citealt{van_Ballegooijen_etal_1989}).

\begin{figure*}[!ht]
\centering
\includegraphics[width=.8\textwidth,clip=]{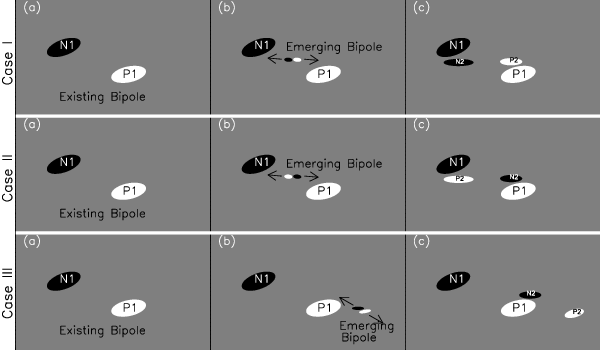}
\caption{Cartoon models showing the changes in the magnetic configuration due to new magnetic flux emergence. The top panels (Case I) show the case of interaction between like magnetic fluxes of emerging and existing magnetic poles. The overall magnetic configuration would be simple, and flare productivity would not change in this scenario. The middle (Case II) and lower panels (Case III) show the cases of interaction between unlike magnetic fluxes of emerging and existing magnetic poles. The overall magnetic configuration would be complex and intense flares would originate in this scenario.}
\label{fig:models_emergence}
\end{figure*}

Certain studies argue that the emergence of a highly twisted magnetic flux tube can produce observational features such as long PILs and shearing motion between opposite magnetic fluxes (e.g., \citealt{Tanaka_1991}; ~\citealt{Fan_etal_1999}). In such scenarios, PILs and shearing motions between opposite magnetic fluxes are observed between conjugate magnetic pairs (magnetic pairs emerging together). This was not the case in our study, as in most cases SgPILs were formed between the nonconjugate pairs (magnetic pairs emerging simultaneously at a different locations or at different times). The emerging opposite magnetic fluxes move away from each other to a certain distance (generally this separation distance depends on the magnetic flux content;~\citealt{Wang_and_Sheeley_1989}). If such a moving pair approaches the magnetic pole (preexisting or emerging simultaneously) of similar polarity, then long SgPILs cannot be formed (case I in Figure \ref{fig:models_emergence}). On the other hand, if such a pair approaches the magnetic pole of opposite polarity, then SgPILs are formed between nonconjugate pairs (see case II and case III in Figure \ref{fig:models_emergence}). The separation motion of the conjugate pair essentially acts as a driver of shearing motion between nonconjugate pairs and the period of interaction between opposite fluxes of nonconjugate pairs depends on the sizes of the emerging pair. If each interacting opposite magnetic polarities (of non-conjugate pair) are of considerable sizes, then prominent SgPILs are observed for a long period of time. Consequently, shearing motion and flux cancellation of opposite magnetic fluxes occur for a long period, and such continuation stores magnetic energy in the solar corona for repetitive flares around the same location, which was observed for SAARs. \cite{Chintzoglou_etal_2019} studied two SAARs in detail. They showed that the convergence of the opposite polarity of nonconjugate pair sresults in flux cancellation along the PILs. They termed the shearing and flux cancellation between the nonconjugate pairs as collisional shearing to differentiate it from the classical cancellation picture of magnetic fluxes between conjugate pairs. This work extended the earlier study of \cite{Chintzoglou_etal_2019} to have more samples to examine the significance of the important conclusion on collisional shearing. This study finds the collisional shearing to be associated with multiple intense flares in almost every selected SAAR. Moreover, our study includes contrasting cases where new magnetic emergence does not result in the interaction between nonconjugate pairs. In such cases, the flare activity does not change either during the emergence or decay phase of ARs. This further supports the idea that collisional shearing is a physical process responsible for multiple intense flares from SAARs. 

The second major objective of this study was the quantification of flare-producing indicators, the so-called feature parameters. Here, we calculated the USF, total unsigned current ($I_{c}$), current helicity ($H_{c}$), degree of current neutralization, length of SgPIL, and $R$ value to analyze the flare productivity of the selected ARs. As FI reflects the time-averaged flare productivity of an individual AR over the period of its disk transit, we compared FI with the average values of magnetic feature parameters of the region. In our study, USF, $I_{c}$, and $H_{c}$ have a weak correlation ($\le$ 0.5) with FI. These magnetic parameters were calculated over the entire AR and reflect the physical processes occurring over the entire region. The value of such parameters depends on the size of ARs and such parameters are considered as extensive properties of ARs (\citealt{Welsch_etal_2009}). $|$DC/RC$|$ has a moderate correlation ($\sim$0.7) with FI. $|$DC/RC$|$ higher than one reflects deviation from current neutrality in AR. \cite{Georgoulis_etal_2012} suggest that nonneutralized current mostly exists around PILs. Deviation from the current neutrality arises due to shearing motion along PILs (\citealt{Torok_2014_EleCurr}; \citealt{Dalmasse_etal_2015}). SgPIL and the $R$ value have the strongest correlation the ($\sim$0.8) with FI. The last three parameters were also calculated over the entire AR and are extensive properties. However, they are related to a subregion of the entire AR, where opposite magnetic fluxes are close to each other. The $R$ value and SgPIL reflect the subregion within the entire AR where shearing motion and flux cancellation can occur. The nonneutralized current is only sensitive to shearing motions. This may be the reason for its weaker correlation than those of the $R$ value and SgPIL. Several studies support the positive relation between SgPIL and intense flares (e.g., \citealt{Welsch_etal_2009}; \citealt{Toriumi_etal_2017}). Our study shows that the convergence of opposite polarity of nonconjugate pairs of considerable size results in the formation of  SgPILs in ARs for a long period. Interaction between opposite magnetic fluxes is an important factor determining the flare productivity of ARs. Our study suggests that the quantification of the processes like shearing motion and flux cancellation along the PIL is important for flare prediction.  

In this study, we only used the occurrence of solar flares to define the energetic state of an AR. However, it shall be noted that both flares and CMEs release stored magnetic energy in the corona and are related to the energetic  state. The standard model considers them as a different manifestation of a single energy release process (solar eruption). There is not always a one-to-one correspondence between flares and CMEs. The association between them increases with the intensity of flares (\citealt{Yashiro_etal_2005}). Therefore, we believe that the number and intensity of flares (especially intense flares) are good for quantifying an AR's energetic state. There are many proposed processes, known as triggering mechanisms, that can initiate the release of stored magnetic energy from ARs (see \citealt{Green_etal_2018}). Here, we used FI to determine the average energetic state of ARs. As we only examined the time-average state of individual ARs, this study is limited in knowing the specific trigger before each flare. Nevertheless, this study is very useful for understanding the dominant processes responsible for long-term magnetic energy buildup in AR. This study shows that most of the intense flares occurred around SgPIL, where shearing motion and flux cancellation occurred over a long period. This result suggests that shearing and cancellation motions are the major energy buildup processes for the intense flares in SAARs.

Finally, we want to comment on the weak correlation between USF and FI for the selected ARs. The very weak correlation in our study is mainly due to the selection effect of these 20 ARs, which are widely distributed in the FI-sunspot size map. It is understood that some other studies show a better correlation between the total flux and intense flare/ flare productivity (e.g., \citealt{Bobra_and_Couvidat_2015}; \citealt{Ran_etal_2022}).  The strong correlation between FI and SgPIL suggests that the area and the period of interaction between opposite magnetic fluxes are important in determining the flare productivity of ARs.  New flux emergence is common around the existing magnetic fluxes (see \citealt{van_Driel_and_Green_2015} and references therein). Therefore, larger ARs (with longer decay periods) would have a higher probability of interaction between opposite fluxes than smaller ARs (with shorter decay periods). This could be the reason for the better relation between the total flux and flares in general. We summarize our major findings in the following list:

\begin{enumerate}
\item  SAARs have complex magnetic configurations compared to MAARs. SAARs have opposite magnetic fluxes in close proximity to each other, generally in the middle of the ARs, giving a complex magnetic configuration and long PILs to ARs. On the other hand, opposite magnetic fluxes are relatively well separated from each other in MAARs, like in a simple bipole with $\beta$-configuration. 
\item Magnetic flux emergence does not necessarily lead to the formation of a complex magnetic configuration. New magnetic flux emergence was observed in both SAARs and MAARs. The interaction between opposite magnetic fluxes of nonconjugate pairs (existing and emerging) created the complex magnetic configuration in SAARs. In MAARs, the interaction between like magnetic fluxes of nonconjugate pairs keeps the overall magnetic configuration simple. 
\item The convergence of opposite polarities of nonconjugate pairs sets up a long-term shearing motion and flux cancellation along the PIL between nonconjugate pairs. Our study found that many of the intense flares from SAARs are located around the PIL. This suggests that multiple intense flares were caused by the persistent shearing motion and flux cancellation.
\item This study found a weaker correlation ($<$ 0.5) between flare productivity (FI) and total flux content, current density, and current helicity. The correlation between flare productivity and the total flux content was the weakest (0.14) for the selected ARs.
\item Our study found a stronger correlation ($>$ 0.5) between flare productivity (FI) and the degree of current neutralization, length of SgPIL, and $R$ value. The correlation between flare productivity and SgPIL/ $R$ value was the strongest ($\sim$0.8). SAARs have SgPIL values longer than 50 Mm or $R$ value greater than 4.5.
\end{enumerate}

\acknowledgments

We thank the referee for the careful reading and for providing constructive comments and suggestions to improve the presentation of the results. We acknowledge the use of data from SDO, in particular, data from HMI. S.Dd and J.Z. acknowledge support from NASA's HTMS program (award No. 80NSSC20K1274) and from the STEREO mission.  S.D. also acknowledges the support of the DKIST Ambassador program. Funding for the DKIST Ambassador program is provided by the National Solar Observatory, a facility of the National Science Foundation, operated under Cooperative Support Agreement No. AST-1400450. J.Z. also acknowledges the support from NASA's SWR2O2R program (award No. 80NSSC23K0451), NASA's MDRAIT program (award No. 80NSSC23K1023), and from NSF multimessenger astronomy supplement funding to NSO (award No. AST-0946422).

\bibliography{flare_productivity}

\begin{thebibliography}{}
\expandafter\ifx\csname natexlab\endcsname\relax\def\natexlab#1{#1}\fi
\providecommand{\url}[1]{\href{#1}{#1}}
\providecommand{\dodoi}[1]{doi:~\href{http://doi.org/#1}{\nolinkurl{#1}}}
\providecommand{\doeprint}[1]{\href{http://ascl.net/#1}{\nolinkurl{http://ascl.net/#1}}}
\providecommand{\doarXiv}[1]{\href{https://arxiv.org/abs/#1}{\nolinkurl{https://arxiv.org/abs/#1}}}

\bibitem[{{Abramenko}(2005)}]{Abramenko_2005}
{Abramenko}, V.~I. 2005, \apj, 629, 1141, \dodoi{10.1086/431732}

\bibitem[{{Antalova}(1996)}]{Antalova_1996}
{Antalova}, A. 1996, Contributions of the Astronomical Observatory Skalnate
  Pleso, 26, 98

\bibitem[{{Bao} {et~al.}(1999){Bao}, {Zhang}, {Ai}, \& {Zhang}}]{Bao_etal_1999}
{Bao}, S.~D., {Zhang}, H.~Q., {Ai}, G.~X., \& {Zhang}, M. 1999, \aaps, 139,
  311, \dodoi{10.1051/aas:1999396}

\bibitem[{{Berger}(2005)}]{Berger_2005}
{Berger}, M.~A. 2005, Highlights of Astronomy, 13, 85

\bibitem[{{Berger} \& {Field}(1984)}]{Berger_and_Field_1984}
{Berger}, M.~A., \& {Field}, G.~B. 1984, Journal of Fluid Mechanics, 147, 133,
  \dodoi{10.1017/S0022112084002019}

\bibitem[{{Bobra} \& {Couvidat}(2015)}]{Bobra_and_Couvidat_2015}
{Bobra}, M.~G., \& {Couvidat}, S. 2015, \apj, 798, 135,
  \dodoi{10.1088/0004-637X/798/2/135}

\bibitem[{{Bobra} {et~al.}(2014){Bobra}, {Sun}, {Hoeksema}, {Turmon}, {Liu},
  {Hayashi}, {Barnes}, \& {Leka}}]{Bobra_etal_2014}
{Bobra}, M.~G., {Sun}, X., {Hoeksema}, J.~T., {et~al.} 2014, \solphys, 289,
  3549, \dodoi{10.1007/s11207-014-0529-3}

\bibitem[{{Cheung} {et~al.}(2017){Cheung}, {van Driel-Gesztelyi},
  {Mart{\'\i}nez Pillet}, \& {Thompson}}]{Cheung_etal_2017}
{Cheung}, M.~C.~M., {van Driel-Gesztelyi}, L., {Mart{\'\i}nez Pillet}, V., \&
  {Thompson}, M.~J. 2017, \ssr, 210, 317, \dodoi{10.1007/s11214-016-0259-y}

\bibitem[{{Chintzoglou} {et~al.}(2019){Chintzoglou}, {Zhang}, {Cheung}, \&
  {Kazachenko}}]{Chintzoglou_etal_2019}
{Chintzoglou}, G., {Zhang}, J., {Cheung}, M. C.~M., \& {Kazachenko}, M. 2019,
  \apj, 871, 67, \dodoi{10.3847/1538-4357/aaef30}

\bibitem[{{Choudhary} {et~al.}(2013){Choudhary}, {Gosain}, {Gopalswamy},
  {Manoharan}, {Chandra}, {Uddin}, {Srivastava}, {Yashiro}, {Joshi}, {Kayshap},
  {Dwivedi}, {Mahalakshmi}, {Elamathi}, {Norris}, {Awasthi}, \&
  {Jain}}]{Choudhary_etal_2013}
{Choudhary}, D.~P., {Gosain}, S., {Gopalswamy}, N., {et~al.} 2013, Advances in
  Space Research, 52, 1561, \dodoi{10.1016/j.asr.2013.07.009}

\bibitem[{{Dalmasse} {et~al.}(2015){Dalmasse}, {Aulanier}, {D{\'e}moulin},
  {Kliem}, {T{\"o}r{\"o}k}, \& {Pariat}}]{Dalmasse_etal_2015}
{Dalmasse}, K., {Aulanier}, G., {D{\'e}moulin}, P., {et~al.} 2015, \apj, 810,
  17, \dodoi{10.1088/0004-637X/810/1/17}

\bibitem[{{Dhakal} {et~al.}(2020){Dhakal}, {Zhang}, {Vemareddy}, \&
  {Karna}}]{Dhakal_etal_2020}
{Dhakal}, S.~K., {Zhang}, J., {Vemareddy}, P., \& {Karna}, N. 2020, \apj, 901,
  40, \dodoi{10.3847/1538-4357/abacbc}

\bibitem[{{Fan} {et~al.}(1999){Fan}, {Zweibel}, {Linton}, \&
  {Fisher}}]{Fan_etal_1999}
{Fan}, Y., {Zweibel}, E.~G., {Linton}, M.~G., \& {Fisher}, G.~H. 1999, \apj,
  521, 460, \dodoi{10.1086/307533}

\bibitem[{{Fletcher} {et~al.}(2011){Fletcher}, {Dennis}, {Hudson}, {Krucker},
  {Phillips}, {Veronig}, {Battaglia}, {Bone}, {Caspi}, {Chen}, {Gallagher},
  {Grigis}, {Ji}, {Liu}, {Milligan}, \& {Temmer}}]{Fletcher_etal_2011}
{Fletcher}, L., {Dennis}, B.~R., {Hudson}, H.~S., {et~al.} 2011, \ssr, 159, 19,
  \dodoi{10.1007/s11214-010-9701-8}

\bibitem[{{Georgoulis} {et~al.}(2012){Georgoulis}, {Titov}, \&
  {Miki{\'c}}}]{Georgoulis_etal_2012}
{Georgoulis}, M.~K., {Titov}, V.~S., \& {Miki{\'c}}, Z. 2012, \apj, 761, 61,
  \dodoi{10.1088/0004-637X/761/1/61}

\bibitem[{{Green} {et~al.}(2011){Green}, {Kliem}, \&
  {Wallace}}]{Green_etal_2011}
{Green}, L.~M., {Kliem}, B., \& {Wallace}, A.~J. 2011, \aap, 526, A2,
  \dodoi{10.1051/0004-6361/201015146}

\bibitem[{{Green} {et~al.}(2018){Green}, {T{\"o}r{\"o}k}, {Vr{\v{s}}nak},
  {Manchester}, \& {Veronig}}]{Green_etal_2018}
{Green}, L.~M., {T{\"o}r{\"o}k}, T., {Vr{\v{s}}nak}, B., {Manchester}, W., \&
  {Veronig}, A. 2018, \ssr, 214, 46, \dodoi{10.1007/s11214-017-0462-5}

\bibitem[{{Hoeksema} {et~al.}(2014){Hoeksema}, {Liu}, {Hayashi}, {Sun},
  {Schou}, {Couvidat}, {Norton}, {Bobra}, {Centeno}, {Leka}, {Barnes}, \&
  {Turmon}}]{Hoeksema_etal_2014}
{Hoeksema}, J.~T., {Liu}, Y., {Hayashi}, K., {et~al.} 2014, \solphys, 289,
  3483, \dodoi{10.1007/s11207-014-0516-8}

\bibitem[{{Kontogiannis} {et~al.}(2017){Kontogiannis}, {Georgoulis}, {Park}, \&
  {Guerra}}]{Kontogiannis_etal_2017}
{Kontogiannis}, I., {Georgoulis}, M.~K., {Park}, S.-H., \& {Guerra}, J.~A.
  2017, \solphys, 292, 159, \dodoi{10.1007/s11207-017-1185-1}

\bibitem[{{Leka} \& {Barnes}(2003)}]{Leka_and_Barnes_2003a}
{Leka}, K.~D., \& {Barnes}, G. 2003, \apj, 595, 1277, \dodoi{10.1086/377511}

\bibitem[{{Li} {et~al.}(2021){Li}, {Chen}, {Hou}, {Veronig}, {Yang}, \&
  {Zhang}}]{Li_etal_2021}
{Li}, T., {Chen}, A., {Hou}, Y., {et~al.} 2021, \apjl, 917, L29,
  \dodoi{10.3847/2041-8213/ac1a15}

\bibitem[{{Liu} {et~al.}(2021){Liu}, {Wang}, {Zhou}, \& {Cui}}]{Liu_etal_2021}
{Liu}, L., {Wang}, Y., {Zhou}, Z., \& {Cui}, J. 2021, \apj, 909, 142,
  \dodoi{10.3847/1538-4357/abde37}

\bibitem[{{Liu} {et~al.}(2017){Liu}, {Sun}, {T{\"o}r{\"o}k}, {Titov}, \&
  {Leake}}]{Liu_etal_2017}
{Liu}, Y., {Sun}, X., {T{\"o}r{\"o}k}, T., {Titov}, V.~S., \& {Leake}, J.~E.
  2017, \apjl, 846, L6, \dodoi{10.3847/2041-8213/aa861e}

\bibitem[{{Parker}(1996)}]{Parker_1996}
{Parker}, E.~N. 1996, \apj, 471, 485, \dodoi{10.1086/177983}

\bibitem[{{Patty} \& {Hagyard}(1986)}]{Patty_and_Hagyard_1986}
{Patty}, S.~R., \& {Hagyard}, M.~J. 1986, \solphys, 103, 111,
  \dodoi{10.1007/BF00154862}

\bibitem[{{Pesnell} {et~al.}(2012){Pesnell}, {Thompson}, \&
  {Chamberlin}}]{Pesnell_etal_2012}
{Pesnell}, W.~D., {Thompson}, B.~J., \& {Chamberlin}, P.~C. 2012, \solphys,
  275, 3, \dodoi{10.1007/s11207-011-9841-3}

\bibitem[{{Ran} {et~al.}(2022){Ran}, {Liu}, {Guo}, \& {Wang}}]{Ran_etal_2022}
{Ran}, H., {Liu}, Y.~D., {Guo}, Y., \& {Wang}, R. 2022, \apj, 937, 43,
  \dodoi{10.3847/1538-4357/ac80fa}

\bibitem[{{Schou} {et~al.}(2012){Schou}, {Scherrer}, {Bush}, {Wachter},
  {Couvidat}, {Rabello-Soares}, {Bogart}, {Hoeksema}, {Liu}, {Duvall}, {Akin},
  {Allard}, {Miles}, {Rairden}, {Shine}, {Tarbell}, {Title}, {Wolfson},
  {Elmore}, {Norton}, \& {Tomczyk}}]{Schou_etal_2012}
{Schou}, J., {Scherrer}, P.~H., {Bush}, R.~I., {et~al.} 2012, \solphys, 275,
  229, \dodoi{10.1007/s11207-011-9842-2}

\bibitem[{{Schrijver}(2007)}]{Schrijver_2007}
{Schrijver}, C.~J. 2007, \apjl, 655, L117, \dodoi{10.1086/511857}

\bibitem[{{Sheeley}(1966)}]{Sheeley_1966}
{Sheeley}, N.~R., J. 1966, \apj, 144, 723, \dodoi{10.1086/148651}

\bibitem[{{Shi} \& {Wang}(1994)}]{Shi_and_Wang_1994}
{Shi}, Z., \& {Wang}, J. 1994, \solphys, 149, 105, \dodoi{10.1007/BF00645181}

\bibitem[{{Shibata} \& {Magara}(2011)}]{Shibata_and_Magara_2011}
{Shibata}, K., \& {Magara}, T. 2011, Living Reviews in Solar Physics, 8, 6,
  \dodoi{10.12942/lrsp-2011-6}

\bibitem[{{Tanaka}(1991)}]{Tanaka_1991}
{Tanaka}, K. 1991, \solphys, 136, 133, \dodoi{10.1007/BF00151700}

\bibitem[{{Tian} {et~al.}(2002){Tian}, {Liu}, \& {Wang}}]{Tian_etal_2002}
{Tian}, L., {Liu}, Y., \& {Wang}, J. 2002, \solphys, 209, 361,
  \dodoi{10.1023/A:1021270202680}

\bibitem[{{Toriumi} {et~al.}(2017){Toriumi}, {Schrijver}, {Harra}, {Hudson}, \&
  {Nagashima}}]{Toriumi_etal_2017}
{Toriumi}, S., {Schrijver}, C.~J., {Harra}, L.~K., {Hudson}, H., \&
  {Nagashima}, K. 2017, \apj, 834, 56, \dodoi{10.3847/1538-4357/834/1/56}

\bibitem[{{T{\"o}r{\"o}k} {et~al.}(2014){T{\"o}r{\"o}k}, {Leake}, {Titov},
  {Archontis}, {Miki{\'c}}, {Linton}, {Dalmasse}, {Aulanier}, \&
  {Kliem}}]{Torok_2014_EleCurr}
{T{\"o}r{\"o}k}, T., {Leake}, J.~E., {Titov}, V.~S., {et~al.} 2014, \apjl, 782,
  L10, \dodoi{10.1088/2041-8205/782/1/L10}

\bibitem[{{van Ballegooijen} \& {Martens}(1989)}]{van_Ballegooijen_etal_1989}
{van Ballegooijen}, A.~A., \& {Martens}, P.~C.~H. 1989, \apj, 343, 971,
  \dodoi{10.1086/167766}

\bibitem[{{van Driel-Gesztelyi} \& {Green}(2015)}]{van_Driel_and_Green_2015}
{van Driel-Gesztelyi}, L., \& {Green}, L.~M. 2015, Living Reviews in Solar
  Physics, 12, 1, \dodoi{10.1007/lrsp-2015-1}

\bibitem[{{Vemareddy}(2019)}]{Vemareddy_2019}
{Vemareddy}, P. 2019, \mnras, 486, 4936, \dodoi{10.1093/mnras/stz1020}

\bibitem[{{Vemareddy} {et~al.}(2015){Vemareddy}, {Venkatakrishnan}, \&
  {Karthikreddy}}]{Vemareddy_2015_flx_emer}
{Vemareddy}, P., {Venkatakrishnan}, P., \& {Karthikreddy}, S. 2015, Research in
  Astronomy and Astrophysics, 15, 1547, \dodoi{10.1088/1674-4527/15/9/011}

\bibitem[{{Wang} \& {Sheeley}(1989)}]{Wang_and_Sheeley_1989}
{Wang}, Y.~M., \& {Sheeley}, N.~R., J. 1989, \solphys, 124, 81,
  \dodoi{10.1007/BF00146521}

\bibitem[{{Welsch} {et~al.}(2009){Welsch}, {Li}, {Schuck}, \&
  {Fisher}}]{Welsch_etal_2009}
{Welsch}, B.~T., {Li}, Y., {Schuck}, P.~W., \& {Fisher}, G.~H. 2009, \apj, 705,
  821, \dodoi{10.1088/0004-637X/705/1/821}

\bibitem[{{Yang} {et~al.}(2017){Yang}, {Hsieh}, {Yu}, \&
  {Chen}}]{Yang_etal_2017}
{Yang}, Y.-H., {Hsieh}, M.-S., {Yu}, H.-S., \& {Chen}, P.~F. 2017, \apj, 834,
  150, \dodoi{10.3847/1538-4357/834/2/150}

\bibitem[{{Yashiro} {et~al.}(2005){Yashiro}, {Gopalswamy}, {Akiyama},
  {Michalek}, \& {Howard}}]{Yashiro_etal_2005}
{Yashiro}, S., {Gopalswamy}, N., {Akiyama}, S., {Michalek}, G., \& {Howard},
  R.~A. 2005, Journal of Geophysical Research (Space Physics), 110, A12S05,
  \dodoi{10.1029/2005JA011151}

\bibitem[{{Zirin} \& {Wang}(1993)}]{Zirin_and_Wang_1993}
{Zirin}, H., \& {Wang}, H. 1993, \nat, 363, 426, \dodoi{10.1038/363426a0}

\end{thebibliography}
\bibliographystyle{aasjournal}

\end{document}